\documentclass[9pt]{IEEEtran}

\usepackage{cite}
\usepackage{amsmath,amssymb}
\usepackage{graphicx,subfig}
\usepackage{hyperref}

\hypersetup{
  colorlinks   = true, 
  urlcolor     = blue, 
  linkcolor    = blue, 
  citecolor   = red 
}

\usepackage{docmute}
\usepackage[table]{xcolor}

\newenvironment{abbreviations}{\begin{list}{}{}}{\end{list}}

\usepackage{tabularx}
\newcolumntype{L}[1]{>{\raggedright\let\newline\\\arraybackslash\hspace{0pt}}m{#1}}
\newcolumntype{C}[1]{>{\centering\let\newline\\\arraybackslash\hspace{0pt}}m{#1}}
\newcolumntype{R}[1]{>{\raggedleft\let\newline\\\arraybackslash\hspace{0pt}}m{#1}}

\usepackage{MathDefs}

\usepackage{verbatim}
\usepackage{tikz}

\makeatother

\begin{document}
\title{Tracking the Progression of Reading Through Eye-Gaze Measurements
\thanks{Submitted for International Conference on Information Fusion, 2019.}
}


\author{
\IEEEauthorblockN{\textbf{Stephen Bottos and Balakumar Balasingam}} \\
\IEEEauthorblockA{
Department of Electrical and Computer Engineering\\
University of Windsor\\
401 Sunset Avenue \\
Windsor, ON, N9B 3P4, Canada\\
\{bottos, singam\}@uwindsor.ca\\
}
}

\maketitle

\begin{abstract}
In this paper we consider the problem of tracking the progression of reading through eye-gaze measurements. 
Such an algorithm is novel and will ultimately help to develop a method of analyzing eye-gaze data which had been collected during reading activity in order to uncover crucial information regarding the individual's interest level and quality of experience while reading a passage of text or book. 
Additionally, such an approach will serve as a ``visual signature'' --- a method of verifying if an individual has indeed given adequate attention to critical text-based information.
Further, an accurate ``reading-progression-tracker'' has potential applications in educational institutions, e-readers and parenting solutions. 
Tracking the progression of reading remains a challenging problem due to the fact that eye-gaze movements are highly noisy and the eye-gaze is easily distracted in a limited space, like an e-book.  
In a prior work, we proposed an approach to analyze eye-gaze fixation points collected while reading a page of text in order to classify each measurement to a line of text; this approach did not consider tracking the progression of reading along the line of text. 
In this paper, we extend the capabilities of the previous algorithm in order to accurately track the progression of reading along each line. 
the proposed approach employs least squares batch estimation in order to estimate three states of the horizontal saccade: position, velocity and acceleration.  
First, the proposed approach is objectively evaluated on a simulated eye-gaze dataset. Then, the proposed algorithm is demonstrated on real data collected by a Gazepoint eye-tracker while the subject is reading several pages from an electronic book. 
\end{abstract}




\section{Introduction}

``The eyes are a window to the soul'' - a commonly spoken expression and one whose relevance increases exponentially as technology advances.  Following the pioneering research of Louis Javal's, psychologist Edmund Huey began to explore the relationship between the ocular behaviour and cognitive processes of test subjects during the act of reading, using the primitive and invasive technology of the early 1900's \cite{huey1908psychology}. Huey's work marked what has come to be known as the first era of eye movement research in the context of cognitive process investigation and behavioural inference. Since, we have progressed through the second era to the third, as documented by Keith Rayner in 1998 \cite{rayner1998eye} in a lengthy collection of works and studies relevant to the field of eye movement tracking during reading and other information processing acts. It was suggested by Rayner that the third era, with its focus on applied, experimental psychology, had been characterized by advances in technology and a rapidly growing interest in the field. At present day, over twenty years later, it may be that we have advanced from the third era to the fourth due to similar reasons. Affordable and advanced eye-tracking technology such as the Tobii \cite{tobii} and Gazepoint \cite{gazept} replace their invasive and expensive counterparts of old. Additionally, interest in the potential to decode raw eye-gaze data in order to infer the otherwise hidden cognitive state of the beholder has never been higher – as said advances in technology have provided the means to perform complex experiments with unprecedented accuracy, similar to the manner by which the processing power of modern-day computers has ushered in the age of Big Data \cite{robert2014machine} by improving the capabilities of Machine Learning as a whole \cite{ friedman2001elements}. 

The study of eye-gaze fixation patterns during various focus-intensive activities has been prevalent for over a century. However, algorithms and systems which grant the capability to accurately track and extract valuable data from clusters of fixation points remain a rarity. Even speaking in terms of state-of-the-art hardware,  inherent inaccuracies introduced  as a result of classification errors by the gaze tracker's computer algorithms \cite{mannaru2017performance, ramdane2008adaptive} make it difficult to identify the precise focal point of visual attention\cite{hyrskykari2006utilizing}. In fact, even if we choose to consider the hypothetical case of one-hundred percent accurate hardware, it is an intrinsic characteristic of the human eye that it rarely stays focused on one point regardless of where the mind is concentrated \cite{istance2008snap,velichkovsky1997towards}. The demand for such algorithms and systems, which are capable of inferring a user’s particular focal points or areas of interest amongst noisy data, exists in both academia and industry, and they have a number of applications. 

One such domain where the demand for accurate eye-gaze tracking techniques has emerged is in the research and development of more effective content recommender systems used by online marketplaces such as Amazon \cite{amazon}, and entertainment platforms such as Netflix \cite{nflix}. Indeed, the placement and delivery of content itself may be optimized with the use of reliable focal point information collected from users \cite{granka2004eye, xu2008personalized, puolamaki2005combining} by examining content which is confirmed by eye-gaze data to have been seen and considered by a user and not acted upon (ie: not clicked), in addition to content which has been acted upon \cite{zhao2016gaze}. Advances in computer vision have attempted to artificially mimic and predict the behaviour of human users \cite{goferman2012context, cerf2009faces}, however a human candidate remains a requirement in order to valiadate their accuracies, and to serve as a benchmark for computer models. As a further constraint, this human produced eye-gaze data must be collected under strictly controlled laboratory conditions \cite{cerf2009faces} which may hinder progression. The idea of using immediate feedback from humans directly as they interact with media has also been proposed \cite{felfernig2013toward}, rather than the mimicking of human behaviour through means of computer vision models.

A particularly interesting subset of content recommender systems are those which suggest educational content, which are relevant to e-learning platforms, otherwise known as Massive Open Online Courses (or MOOCs), such as Coursera \cite{csera}. It is imperative, for such platforms to succeed, to generate personalized recommendations based on individual user interest and background knowledge \cite{vaishali2016learning, tang2005smart}. In addition to estimating interest based on engagement (ie: negative ratings \cite{albatayneh2018utilizing}), content presented by MOOC recommendation algorithms also incorporate more intimate, learner-specific factors and seek to investigate exactly how an individual prefers to educate themselves \cite{syed2018personalized, daradoumis2013review, zaiane2001web}. On a related note, research has attempted to quantify a reader's interest in various passages of text \cite{wade1999using, babbitt2004assessing} in order to determine exactly what criteria must be included to appeal to that individual - particulary, the presence of information which was important to the reader, connections made by the reader between the text and their own personal experiences, and adequate background information.

In this paper, we continue the development of previous work \cite{bottos2019approach}, which aims to produce a reliable method of processing raw eye-gaze fixation point data collected during reading in order to extract valuable insights regarding an individual's cognitive state - interest level, or comprehension for example - as well as to verify exactly which lines of text were read and which were not. This technology is intended to prove useful in conjunction with the previously discussed applications. For example, in the case of recommender systems, one can imagine the advantages of employing crowdsourcing platforms such as Amazon's Mechanical Turk \cite{mturk} to collect validation data for computer vision algorithms rather than relying on highly controlled laboratory environments, with the help of application-specific algorithms to reduce inherent noise and grant a simple webcam the ability to produce meaningful results. In a similar vein, rather than relying on content engagement (ie: clicks, negative votes), one may analyze content that has been recommended to an individual, but which has been examined and ignored. In the special case of MOOCs, this can prove invaluable as course syllabi are typically text-based media. On the topic of MOOCs, the ability to analyze reading patterns will also grant insights as to how the individual prefers to learn and consume content.

\section{Problem Definition} \label{sec:problem_def}
\label{probdefs}

  \begin{figure*}[h]
  \begin{center}
  \subfloat[]{\includegraphics[width=.32\textwidth]{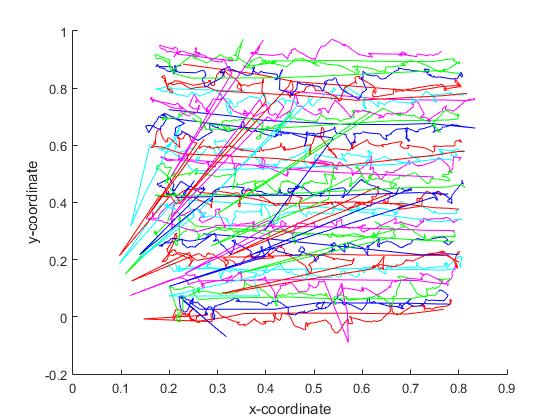} \label{a} } \hspace{1pt}
  \subfloat[]{\includegraphics[width=.32\textwidth]{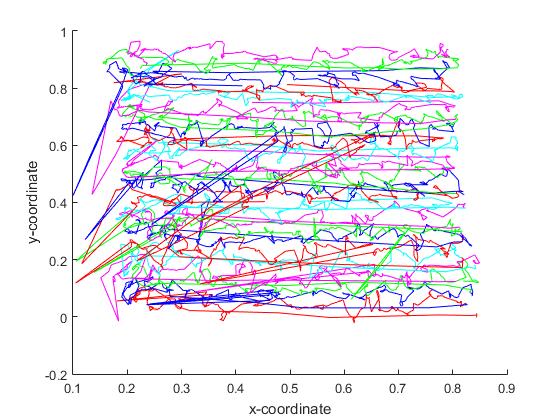} \label{b} }
\hspace{1pt}
  \subfloat[]{\includegraphics[width=.32\textwidth]{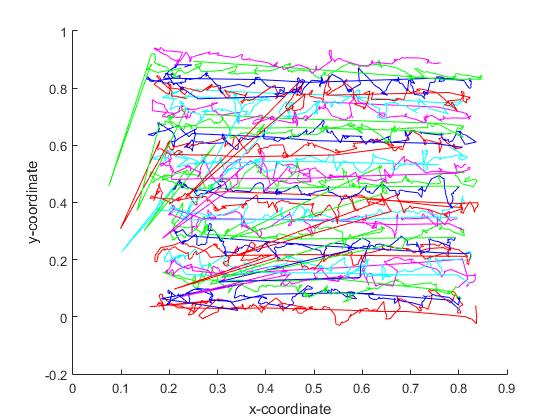} \label{c} }
  \caption{{\bf Real eye-gaze fixation points, colour-coded per line.} Real eye-gaze fixation points collected during the reading of three seperate 25-line passages of text are shown, in order to showcase the presence of ``unwanted backtracks'', both subtle and extreme, present in the data.
  }
  \label{backtracks}
  \end{center}
  \end{figure*}  

\def\midas{In reference to the Greek mythological King who got his wish - to turn everything that he touched into gold - granted. }

The main goal of our past and current work is to develop a system which grants the ability to detect and verify a reader's progression through a block of text using artificial means, even when input data is highly corrupted by noise. Firstly, let us define natural reading progression. If one is to consider a page in a novel, containing multiple lines of text of equal length and equal vertical spacing, then natural reading progression, speaking in terms of non-Semitic languages, will begin at the top-left of the page and track left to right. When one line of text is completed, the reader will begin the same left-to-right progression on a new line immediately below that which was previously read. Second, let us illustrate what is meant when we describe input data as ``highly corrupted by noise''. Figure \ref{backtracks} demonstrates a few pages worth of real-world data, collected during reading of a 25-line passage of text. Each individual line has been colour-coded for convenience. One can observe that data-points belonging to individual lines exhibit erratic behaviour, often overlapping rather than appearing distinctly seperable. Indeed, as was previously mentioned, noise is introduced to eye-gaze data from two sources, which are as follows:
\begin{itemize}
\item The {\em measurement noise} introduced  as a result of classification errors made by the gaze tracker's computer algorithms \cite{mannaru2017performance, ramdane2008adaptive}.
\item The {\em MIDAS\footnote{\midas} touch problem} in eye-gaze tracking. It is an intrinsic characteristic of the human eye that it rarely stays focused in one point regardless of where the mind is concentrated \cite{istance2008snap,velichkovsky1997towards}.
\end{itemize}
\par
In this paper, we attempt to improve upon previous work in which the objective was to predict, using eye-gaze fixation points, the true line of text being read at any given instant from some text passage of interest despite the presence of noise. This approach, named the ``Line Detection Algorithm'' (LDS), utilized only the $y$-coordinate component of an individual's eye-gaze fixation point, which described the location between the top and bottom of the page that an individual's gaze was recorded at that instant. The next step is to incorporate the $x$-coordinates into the prediction along with the predicted line.

\par Referring again to Figure \ref{backtracks}, one can observe that erratic behaviour is not only present in the form of individual lines overlapping one another, but also in the form of what we will refer to as \emph{false saccades}. False saccades are defined as quick saccades in any direction that are not a result of natural reading progression, and should be discarded.  Due to the nature of the data collection process, no deviation from natural reading progression should be observed, and yet through examination of Figure \ref{backtracks} one can see that both subtle and extreme false saccades exist.  These false saccades may be the result of noise introduced by the points mentioned above, and are not true representations of the reader's progression through a line of text.

\par Similarly, what may appear to be a false saccade may indeed be the result of a reader skipping ahead through a passage of text, or returning to a previously read portion. This means that if the reader makes a quick saccade from one point in the text to another and begins reading anew from that point, then this should be reflected in the $x$-coordinate prediction rather than ignored. The objective of this paper's research is to develop an approach to accurately track the progression of eye-gaze fixations while reading. We propose a Least Squares filtering technique which will protect against false saccades in order to infer a smooth, more accurate and natural-looking reading progression when coupled with the previously developed LDS than can be predicted using the raw data alone.

\section*{List of Notations I}
\begin{abbreviations}
\item[$\bz_x$] the vector containing each measured $x$-coordinate, where $z_x(i)$ represents a single eye-gaze fixation point in $\bz_x$, $i = 1:T$, where $T$ is the total number of eye-gaze measurements per page
\item[$\bz_y$] the vector containing each measured $y$-coordinate, where $z_y(i)$ represents a single eye-gaze fixation point in $\bz_y$, $i = 1:T$
\item[$\bx$] the vector containing each true $x$-coordinate, where $x(i)$ represents a single eye-gaze fixation point in $\bx$, $i = 1:T$
\item[$\by$] the vector containing each true $y$-coordinate, where $y(i)$ represents a single eye-gaze fixation point in $\by$, $i = 1:T$
\item[$L(i)$] the true line being read at $i$, based on the $y$-coordinate of the eye-gaze fixation at $i$
\item[$\hat{\bL}$] the vector containing all line estimates for each eye-gaze fixation point, where $\hat{L}(i)$ represents the estimated line at $i$
\item[$N$] Number of lines in a particular text of interest 
\item[$\bz_{x,n}$] the vector containing all $x$-coordinates $z_x(i)$ belonging to the estimated line $n$, which satisfy the condition $\hat{L}(i) = n$, where $n \in {1,..., N}$
\item[$\bx_n$] the vector containing each true $x$-coordinate belonging to a specific line
\item[$\hat{\bx}_n$] the estimate of $\bx_n$
\item[$l_n$] the length of the vectors $\bz_{x,n}$, $\bx_n$, and $\hat{\bx}_n$, for $n = 1:N$
\item[$z_{x,n}$] the measured $x$-coordinate of the eye-gaze fixation belonging to line $n$, where $ k = 1:l_n$
\item[$\hat{x}_n(k)$] the estimated true $x$-coordinate of the eye-gaze fixation belonging to a line $n$, where $ k = 1:l_n$
\item[$\hat{\bx}$] the vector containing all estimated $x$-coordinates $\hat{x}(i)$, the restult of each $\hat{\bx}_n$ vector stacked in ascending order per line number
\item[$\Delta T$] the sampling rate at which measurements are obtained, $\Delta T = 64$
\item[${\sigma}$] the standard deviation of the Gaussian noise distribution
\item[$v_n(k)$] the measurement noise value at $k$ in batch $n$, a single random sample drawn from the distribution $\mathcal{N}(0,\sigma)$
\item[${\bF(k)}$] the state model which describes the system, given as follows:
\begin{eqnarray}
{\bF(k)} = 
\left[
\begin{array}{ccc} 
1 & k\times \Delta T & \frac{k^2\times \Delta T^2}{2} \\
0 & 1 & k\times \Delta T \\
0 & 0 & 1
\end{array} 
\right]
\end{eqnarray}
\item[${\bf H}_n$] the rows of the state model corresponding to the parameter of interest. In this case we are interested in estimating position only, so,
\begin{eqnarray}
{\bf H}_n = 
\left[
\begin{array}{ccc} 
1 & 1\times \Delta T & \frac{k^2\times \Delta T^2}{2} \\
\vdots & \vdots & \vdots \\
1 & l_n\times \Delta T & \frac{l_n^2\times \Delta T^2}{2} \\
\end{array} 
\right]_{l_n\times 3}
\end{eqnarray}

\item[$\bH_n(k)$] it is often necessary to discuss $\bH_n$ in terms of its individual rows only rather than as a whole, thus let
\begin{eqnarray}
{\bH_n(k)} = 
\left[
\begin{array}{ccc} 
1 & k\times \Delta T & \frac{k^2\times \Delta T^2}{2} \\
\end{array} 
\right]_{l_n\times 3}
\end{eqnarray}

\item[${\bf R_n}$] a diagonal matrix containing the measurement variance as follows
\begin{eqnarray}
{\bf R}_n = \sigma^2 {\bl_{l_n\times l_n}}
\end{eqnarray}

\end{abbreviations}

\section{Proposed Approach}
\subsection{Pre-Processing}

  \begin{figure*}
  \centering
  \includegraphics[width=6.2in]{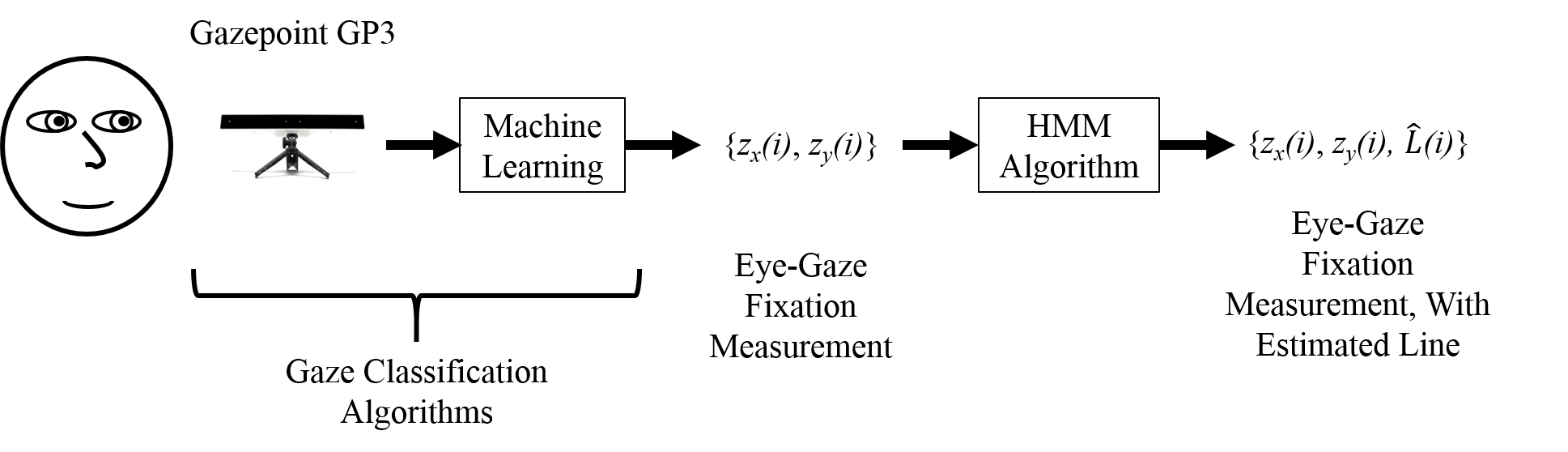}
  \caption{
  {\bf Line detection system (LDS).}
  A block diagram describing the function of the previously developed LDS. 
  }
  \label{generalalg}
  \end{figure*}

\begin{figure}
\begin{center}
\includegraphics[width = .55\columnwidth]{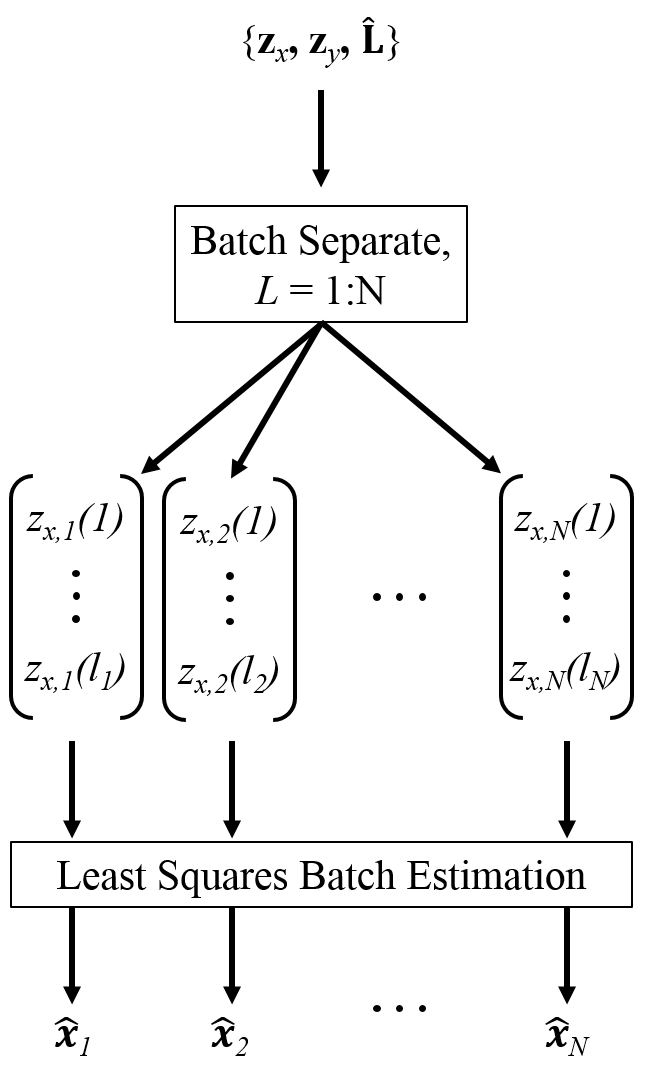}
\caption{{\bf Batch separation.} Separation of the total data set into batches according to their estimated line.}
\label{batchsplit}
\end{center}
\end{figure}

Pre-processing of the data was performed in accordance with the procedure demonstrated in the previous paper \cite{bottos2019approach}. In summary, measured eye-gaze fixation points were fed to an algorithm whose main decision-making process was driven by a Hidden Markov Model (HMM). The input to the algorithm were raw eye gaze fixation coordinates, and the output of the algorithm were the original, unmodified eye-gaze fixation points along with the predicted line of text that the user was reading from at the time corresponding to each fixation point. A graphical representation of the algorithm is given in Figure \ref{generalalg}.

\par For a given area of interest, containing a passage of text having $N$ lines, a full data-set is able to be separated into batches according to the estimated line $\hat{L}(i)$ assigned to each eye-gaze fixation point. This is necessary prior to performing the Least Squares batch estimation step (LS). In accordance to the previously defined naming convention, and assuming that  $n \in {1,2,3,...,N}$, it is possible to develop the set of batches,
\begin{equation*}
\left\{  \underset{l_1 \times 1}{\bz_{x,1}}, \underset{l_2 \times 1}{\bz_{x,2}} , \underset{l_3 \times 1}{\bz_{x,3}},..., \underset{l_N \times 1}{\bz_{x,N}}                            \right\}
\end{equation*}
which contain all $z_x(i)$ points belonging to each estimated line. Each batch of measured $x$-coordinates is now ready to be processed using the LS method in order to obtain its corresponding estimated true $x$-coordinates.  Figure \ref{batchsplit} provides a visualization of how each batch is created from the total set of data. From this point forward we will speak in terms of some arbitrary line $n$. 

\subsection{Least Squares Batch Estimation}
We enter this estimation step with a batch of $x$-coordinates, the measurements $\bz_{x,n}$, belonging to some arbitrary line. Each batch of $x$-coordinates is a subset of the eye-gaze fixation coordinates collected during reading of the passage of text of interest, and thus this process will be repeated for each line of text detected. 
\par
\begin{figure}
\begin{center}
\includegraphics[width = .85\columnwidth]{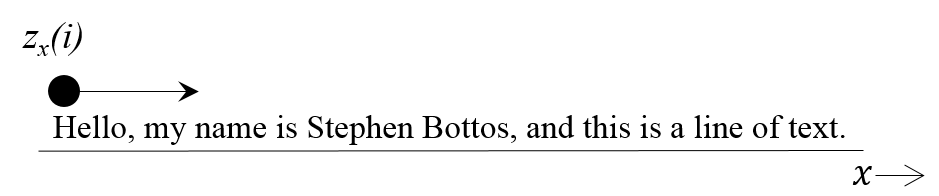}
\caption{{\bf Eye-gaze fixation progression.} While reading, an individual's eye-gaze is assumed to progress from left to right for each line of text read.}
\label{movingob}
\end{center}
\end{figure}

 To re-iterate, the objective of this particular step is to filter out presumably unwanted noise, as was demonstrated in figure \ref{backtracks}, in order to obtain a smooth, less erratic progression in the $x$-direction which is more akin to natural reading. We can imagine an individual's progression through a line of text as if his or her gaze is an object in motion, such as in Figure \ref{movingob}, considering the horizontal component of the eye-gaze fixation point only. 

\par Now, $z_{x,n}(k)$ descibes the measured $x$-position of the eye-gaze fixation point only. It must now be stressed that $z_{x,n}(k)$ is the \emph{measured value} of the $x$-coordinate of the eye-gaze fixation point. There is an element of noise introduced to each measurement by both the instrumentation used to obtain the measurement as well as the natural movement of the human eye as was previously discussed, and external influences such as, in our case, head movement or blinking (among others). Thus, the \emph{estimated value}, $\hat{x}_n(k)$, is contained within the measured value along with some level of corruption due to noise, which will be illustrated mathematically momentarily. To fully describe the behaviour of any moving object we must also include its velocity,  $\dot{x}_n(k)$, and acceleration, $\ddot{x}_n(k)$, components. Together, these three components define the \emph{state at $k$} of the object as a vector - $\bx_n(k)$ in our case - which can be written as follows:

\begin{eqnarray}
\bx_n(k) = \left[ 
\begin{array}{c}
x_n(k)\\
\dot{x}_n(k)\\
\ddot{x}_n(k)
\end{array}
\right]
\end{eqnarray}

\par With this, it is possible to estimate the state of the object at the next time instant by multiplying the state by the model ${\bF(1)}$ as follows,

\begin{eqnarray}
\bx_n(k+1)  = \bF(1) \bx_n(k)
\end{eqnarray}
or,
\begin{eqnarray}
\bx_n(k+1) = 
\left[
\begin{array}{ccc}
1 & \Delta T & \frac{\Delta  T^2}{2} \\
0 & 1 & \Delta  T \\
0 & 0 & 1
\end{array}
\right]
\left[
\begin{array}{c}
x_n(k)\\
\dot{x}_n(k)\\
\ddot{x}_n(k)
\end{array}
\right]
\end{eqnarray}
It is assumed that the duration between the $k^{\rm th}$ and $(k+1)^{\rm th}$ time interval is the previously defined sampling rate, $\Delta  T$, a constant. 

\par  The measured value, $z_{x,n}(k)$ is equivalent to an estimated true value plus some amount of noise corruption. Mathematically, this can be expressed as, 

\begin{eqnarray}
\label{main_eq}
z_{x,n}(k) &=& \left[ 
\begin{array}{ccc}
1 &0 & 0
\end{array}
 \right] \bx_n(k) + v_n(k) \nonumber\\
&=&
\left[ 
\begin{array}{ccc}
1 &0 & 0
\end{array}
 \right]
 \left[ 
\begin{array}{c}
x_n(k)\\
\dot{x}_n(k)\\
\ddot{x}_n(k)
\end{array}
\right] + v_n(k) \nonumber\\
&=& x_n(k) + v_n(k)
\end{eqnarray}

\noindent where only the position component of $\bx(k)$ is extracted, since that is the parameter of interest for this particular problem. 

\par The state of the object at any time step $k$ can be written in terms of the initial state as follows: 
\begin{eqnarray}
\label{init}
\bx_n(0) = \left[ 
\begin{array}{c}
x_n(0)\\
\dot{x}_n(0)\\
\ddot{x}_n(0)
\end{array}
\right]
\end{eqnarray}

\begin{eqnarray}
\bx_n(1) &=& \bF(1) \bx_n(0) \\
\bx_n(2) &=& \bF(2) \bx_n(0) \\
\bx_n(3) &=&\bF(3) \bx_n(0) \\
&\vdots& \\
\bx_n(l_n) &=& \bF(l_n) \bx_n(0)
\end{eqnarray}

\noindent with eq. \ref{init} used to initialize the calculation and where $k = 1:l_n$, the length of the batch. Correspondingly, considering eq. \ref{main_eq} and the fact that the only parameter of interest is position, the measurement equation can be simplified with the substitution of $\bF(k)$ with $\bH_n(k)$ as follows:

\begin{eqnarray}
\label{eq10}
\begin{array}{ccc}
x_n(1) = \bH_n(1)\bx_n(0) + v_n(1)
\\
x_n(2) = \bH_n(2)\bx_n(0) + v_n(2)
\\
x_n(3) = \bH_n(3)\bx_n(0) + v_n(3)
 \\
\vdots 
\\
x_n(l_n) = \bH_n(l_n)\bx_n(0) + v_n(l_n)
\end{array}
\end{eqnarray}

It is possible to vectorize eq \ref{eq10} in the form,

 \begin{eqnarray}
 \label{11}
\bx_n = \bH_n \bx_n(0) + \bv_n
\end{eqnarray}
where,
\begin{eqnarray}
\bx_n &=& [x_n(1) \,\, x_n(2) \,\, x_n(3) \,\, \ldots \,\, x_n(l_n) ]' \nonumber \\
\bH_n &=& 
\left[
\begin{array}{ccc}
1 &\Delta  T & \frac{\Delta  T^2}{2}\\
1 & 2\Delta T & \frac{2^2\Delta T^2}{2} \\
1 & 3\Delta T & \frac{3^2\Delta  T^2}{2} \\
\vdots & \vdots & \vdots \\
1 & l_n\Delta T & \frac{l_n^2\Delta  T^2}{2}\\
\end{array}
\right]\\
\bv_n &=& [v_n(1) \,\, v_n(2) \,\, v_n(3) \,\, \ldots \,\, v_n(l_n) ]'
\end{eqnarray}

\par The key takeaways from the discussion thus far are that the estimated true value is contained within the measured value along with some amount of noise corruption, and that by initializing the estimated true state $\hat{\bx}_n(0)$ it is possible to estimate all other $x$-positions $\hat{\bx}$ in a given batch $n$. The initial estimated state, $\hat{\bx}_n(0)$, is obtained as follows:
\begin{equation}
\hat{\bx}_n(0) = \left[\bH_n'(\bR_n)^{-1}\bH_n\right]\bH_n'(\bR_n)^{-1}\bx_n
\end{equation}
which, accounts for the noise by way of the noise covariance matrix. $\bR_n$. Finally, similarly to eq. \ref{11}, the estimates for the batch corresponding with line $n$ can be obtained as such:
 \begin{eqnarray}
 \label{all_ests}
\hat{\bx}_n = \bH_n \hat{\bx}_n(0)
\end{eqnarray}

\section{Computer Analysis}
\subsection{Simulated Data}

  \begin{figure*}[h]
  \begin{center}
  \subfloat[][$\sigma = 0.2$ line-width]{\includegraphics[width=.3\textwidth]{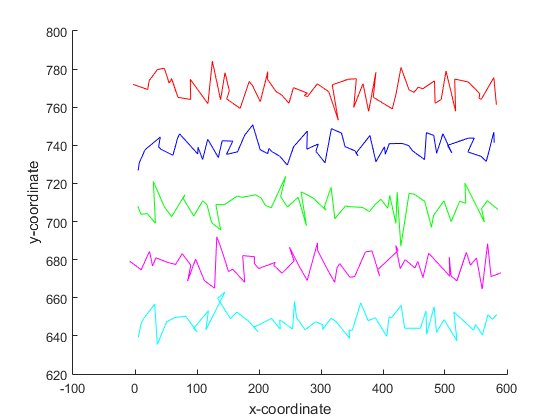} \label{a} } \hspace{1pt}
  \subfloat[][$\sigma = 0.46$ line-width]{\includegraphics[width=.3\textwidth]{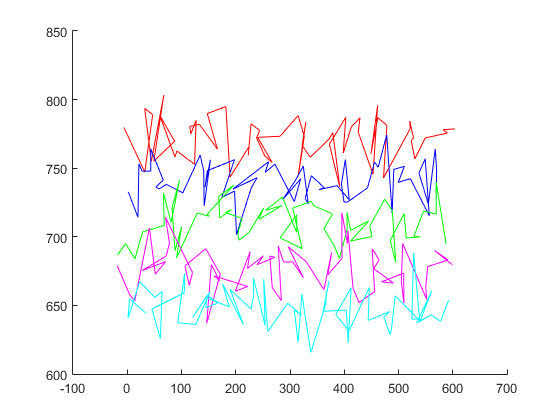} \label{b} }
\hspace{1pt}
  \subfloat[][$\sigma = 1$ line-width]{\includegraphics[width=.3\textwidth]{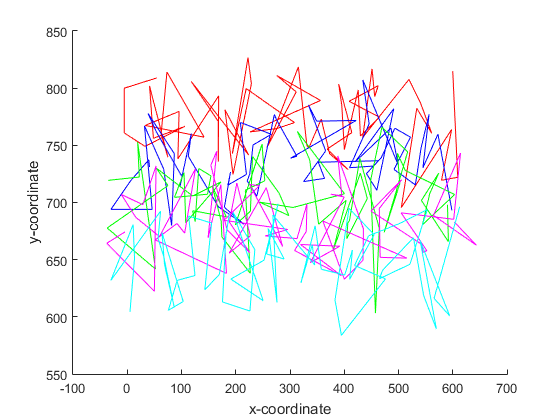} \label{c} }
  \caption{{\bf Simulated eye-gaze measurements for different measurement noise levels.}
  The $(x,y)$-coordinates of the eye-gaze point data while reading lines 1 to 5 are shown for different noise levels. 
  The unit of the measurement noise standard deviation is shown in ``line-widths,'' i.e., the distance between two adjacent lines. 
  }
  \label{simdata}
  \end{center}
  \end{figure*}

\begin{figure}
\begin{center}
\includegraphics[width = .55\columnwidth]{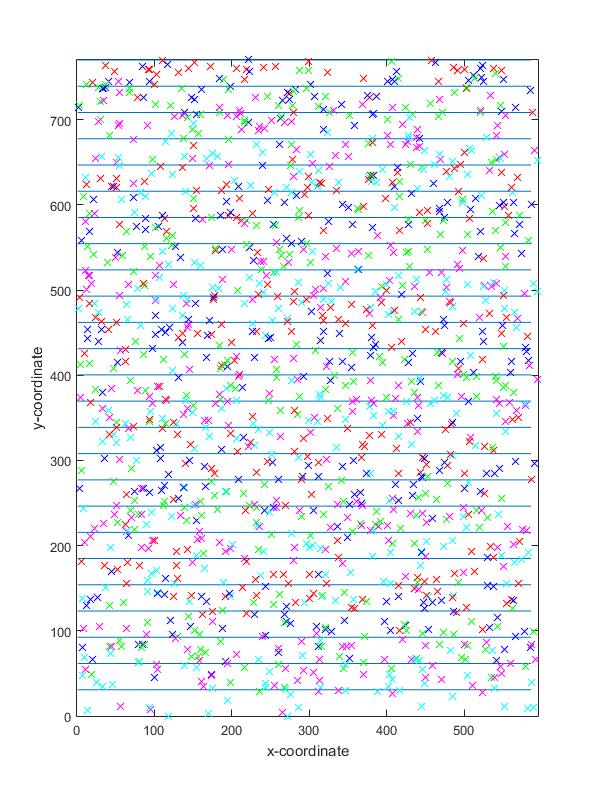}
\caption{{\bf Simulated data, full page} A full ``page'' of simulated data consisting of 25 lines, where data-points belonging to each line have been colour-coded for visibility.}
\label{fullpg}
\end{center}
\end{figure}

Simulated data was produced at varying noise levels, $\sigma$, the standard deviation of the distribution from which each data point was generated where $\sigma_x = \sigma_y$. For the sake of convenience, $\sigma$ is expressed in terms of line widths - the distance between each line of text. For example, an instance where $\sigma = 1$ means that each $x$ and $y$ coordinate will include some random amount of noise from a distribution whose standard deviation is equal to the distance between each line. Likewise, an instance where $\sigma = 0.5$ will generate noise whose standard deviation is equal to half the distance between each line. Nine levels of $\sigma$ were tested, which were as follows, in ascending order:
\begin{equation}
\sigma = \left\{0.2, 0.22, 0.25, 0.26, 0.3, 0.37, 0.46, 0.63, 1\right\}
\end{equation}
\par
For each noise level, 20 ``pages'' of simulated data were created. Each individual page of simulated data contained 25 lines, thus $N$ = 25. In Figure \ref{simdata}, colour coded five-line samples of data generated at $\sigma = 0.2$, $\sigma = 0.46$, and $\sigma = 1$ are shown, while Figure \ref{fullpg} illustrates a full ``page'' worth of simulated data, generated at $\sigma = 0.2$.
  
\subsection{Performance Evaluation Metric, Simulated Data}
The process of obtaining estimated $x$-coordinates has been covered in previous sections. For each noise level, 20 full batches of simulated data were created, which were each divided into smaller batches on a per-line basis using the line estimates from the LDS. The performance of the line-detection algorithm (first parse) is discussed in detailed in \cite{bottos2019approach}. 

By utilizing the batches $\bx_n$ and $\hat{\bx}_n$, for each line, the  horizontal-saccade-tracking algorithm is evaluated by comparing the true $x$-coordinate with the estimated $x$-coordinate, however only for the cases in which the estimated line is also correct. That is, let us say, that for a particular $\hat L_i $, such that  $\hat L_i = L_i$, there are $m$ indices, $j= i_1, i_2, \ldots, i_m$ of the data points corresponding to the estimated line $\hat L_i$. 
For this,  the LS estimated x-points are given by
\begin{eqnarray}
[\hat x(i_1), \hat x(i_2), \ldots, \hat x(i_m)]
\end{eqnarray}

The Normalized Root Mean Square Error (NRMSE) of the horizontal-saccade-tracking algorithm is then defined as, 
\begin{eqnarray}
{\rm NRMSE}_p = 
\left( \sqrt{\frac{ \sum_{j=1}^{m} (\hat x(i_j) - x(i_j))^2 }{m}} \right) \times \frac{100}{600}
\end{eqnarray}
where $p$ denotes the ``page number'', and the normalization term allows the RMSE to be expressed as a percentage of the text-width boundaries chosen within which to generate simulated data. Since 20 pages of simulated data were collected, the average NRMSE was computed as.
\begin{eqnarray}
{\rm NRMSE} = 
\frac{ \sum_{p=1}^{20} {\rm NRMSE}_p }{ 20 }
\end{eqnarray}
\subsection{Results, Simulated Data}
\begin{figure}
\begin{center}
\includegraphics[width = .9\columnwidth]{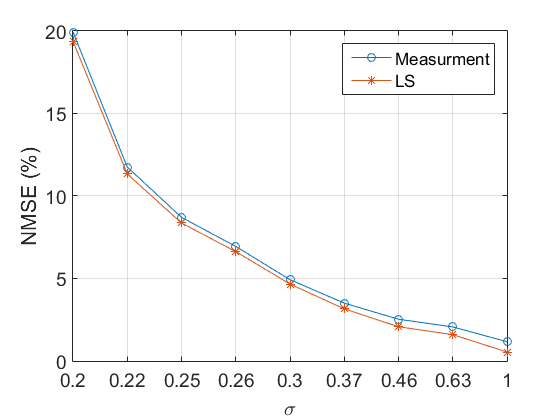}
\caption{{\bf NRMSE plots}. The NRMSE for each noise level at which simulated data was generated is plotted, results are shown for both the measured $x$-coordinates and the estimated $x$-coordintes.}
\label{error}
\end{center}
\end{figure}

\begin{figure}
\begin{center}
\includegraphics[width = .9\columnwidth]{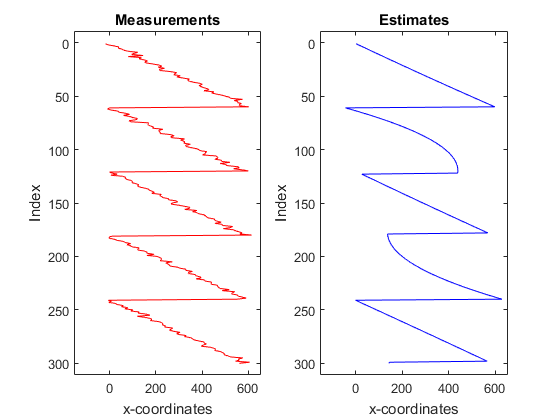}
\caption{{\bf Measured and estimated results comparison.} The smoothing effect of the horizontal-saccade-tracking algorithm is demonstrated by plotting the $x$-coordinates of the measured data alongside the estimated $x$-coordinates. Only the first five lines of results are given, to conserve space. This sample is taken from data produced at $\sigma = 0.46$.}
\label{smoothing}
\end{center}
\end{figure}

The NRMSE between the measured $x$-coordinate and the true $x$-coordinate, and the error between the estimated $x$-coordinate and the true $x$-coordinate at varying levels of $\sigma$ were computed and comparitively plotted in Figure \ref{error}. An improvement, albeit slight, can be observed between the estimated $x$-coordinates using the LS method and their simulated measured counterparts. However - recall that the desire to reduce the presence of \emph{unwanted backtracks} was also discussed previously. The measured and estimated $x$-coordinates (for the first five lines only, to conserve space) were plotted in Figure \ref{smoothing}, for $\sigma = 0.46$. While on a per-line basis the addition of the LS estimator managed to reduce the presence of unwanted noise, this effect can only be deemed meaningful if the line estimate is correct in the first place. It can be observed that in some instances, the line-batch to be processed contains too many or too few predictions, introducing an element of error into the x-estimations.
 
\section{Application of the Proposed Approach to Gazepoint Data}
\begin{figure*}
  \begin{center}
  \subfloat[]{\includegraphics[width=.32\textwidth]{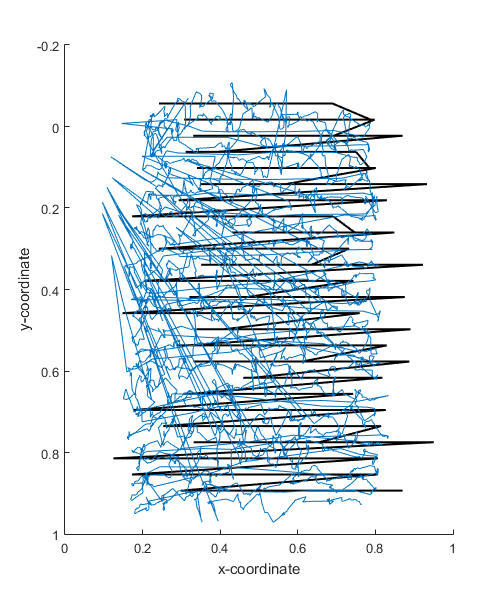} \label{a} } \hspace{1pt}
  \subfloat[]{\includegraphics[width=.32\textwidth]{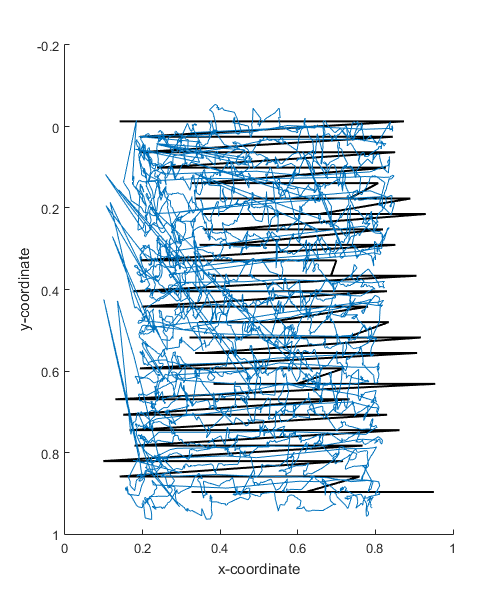} \label{b} }
\hspace{1pt}
  \subfloat[]{\includegraphics[width=.32\textwidth]{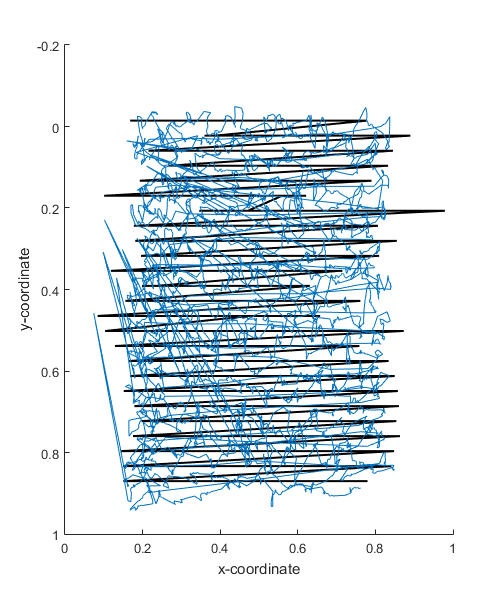} \label{c} }
  \caption{{\bf Estimated $(x,y)$-coordinates, overlayed on top of their corresponding real-data sets.}
  We demonstrate the effectiveness of the LDS and the horizontal-saccade-tracking algorithm on three pages of real-data collected for 25-line pages. The line to which each eye-gaze fixation point belongs is estimated by the LDS, along with its estimated $y$-position on the page. The output of the horizontal-saccade-tracking algorithm is then used to estimate the $x$-coordinates of each point, eliminating false saccades.
  }
  \label{realdata}
  \end{center}
  \end{figure*}

It was previously discussed in Section \ref{probdefs} that the objective of the proposed approach in this paper was to eliminate false saccades and unveilthe true reading pattern hidden amidst noisy data. The result of our efforts is given in Figure \ref{realdata}, which shows the estimated lines and $x$-coordinates overlayed with the raw eye-gaze data, where the $y$-coordinate of each line on the page and the eye-gaze fixations belonging to a particular line were determined by the LDS.
\par
\begin{figure*}
  \begin{center}
  \subfloat[]{\includegraphics[width=.32\textwidth]{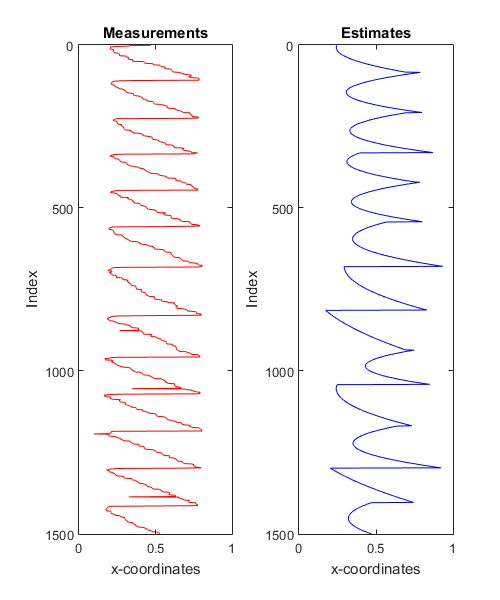} \label{a} } \hspace{1pt}
  \subfloat[]{\includegraphics[width=.32\textwidth]{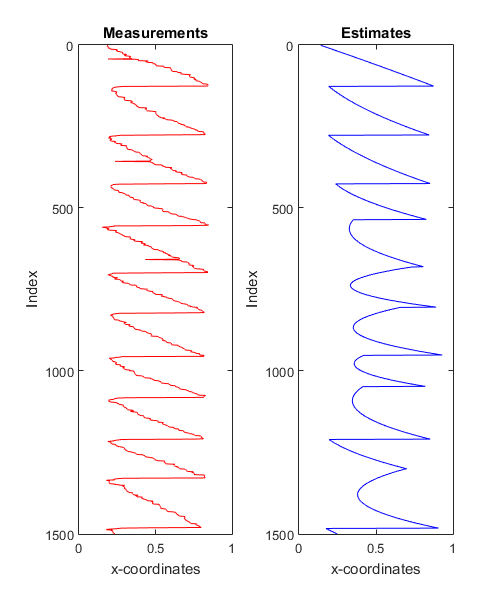} \label{b} }
\hspace{1pt}
  \subfloat[]{\includegraphics[width=.32\textwidth]{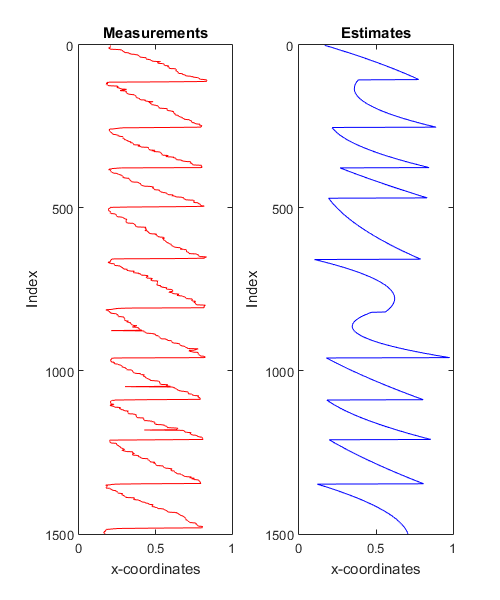} \label{c} }
  \caption{{\bf Measured $x$-coordinates and estimated $x$-coordinates.}
  The first 1500 data points, to conserve space, for both the measured $x$-coodinates from the real data and their corresponding estimated $x$-coordinates are plotted side-by-side in order to visually compare the level of noise in each. 
  }
  \label{realdatasmooth}
  \end{center}
  \end{figure*}

 
Due to the fact that no ground truth $x$-coordinates were available for the real-world data, the following knowledge of the data-collection procedure was used in order to visually examine the aforementioned figures for errors:
\begin{itemize}
\item each of the 25 lines should be read exactly once, in sequence, starting with the first line at the top of the page
\item each line is of roughly equivalent length, and each line is read in full from left to right before advancing to the next
\item no skipping forward or backward within a line was permitted, meaning there should be no false saccades 
\end{itemize}
In essence, the data-collection procedure was designed to mimic natural reading progression. Referring to \ref{realdata}, one can observe that the result of our proposed approach of estimating all $\hat{\bL}$ and $\hat{\bx}$ manages to extract the true reading pattern hidden within the noisy data. Since \ref{realdata} is unable to display the presence of any false saccades within the estimated points, \ref{realdatasmooth} has been included alongside the raw data for comparison - using only the first 1500 eye-gaze fixations from each in order to conserve space. Through examination, one can observe that the LS estimation procedure is highly effective at eliminating false saccades.
\par
While the LS estimation procedure yields positive results, however, its accuracy is dependant on the predictions made by the LDS in the same manner as the simulated data. One can clearly observe cases where the a given line contains too few or too many data points are present for a given line.

\section{Conclusions and Discussions}

In this paper, we presented an approach to track the progression of horizontal reading based on eye-gaze measurements. 
The proposed approach uses a previously developed approach by the authors to detect the line number of each data point; then, an approach based on the least squares method is proposed to estimate the progression of the eye-gaze along horizontal lines.

While results on both simulated and real data show promise in terms of eliminating false saccades using the horizontal-saccade-tracking algorithm, the formulation of each batch to be processed --- that is, the batch containing all eye-gaze fixation points estimated to belong to a certain line by the line detection system (LDS) \cite{bottos2019approach} --- limits the accuracy of this proposed algorithm. By incorrectly estimating the line to which any eye-gaze fixation point belongs, an $x$-coordinate that should not be considered in that line's batch is computed by the horizontal-saccade-tracking algorithm, this was found to be the important cause towards the estimation error that was objectively computed using the simulated data. 

Our immediate future direction of this research will focus on the following two aspects: 
(i) {\em Improvements to the LDS algorithm:} Improved LDS algorithm will input relevant points for the horizontal estimation/tracking. Further, the LDS algorithm in  \cite{bottos2019approach} assumes that the number of lines in a passage of interest is known a-priori. More sophisticated statistical model and information theoretic rules can be employed to improve the existing LDS algorithm.
(ii) {\em Improvements to horizontal saccade tracking algorithm:}
The proposed least squares algorithm assumes a noiseless process model -- this was selected based on the initial inspection of the real data and its tendency to "drift into" wrong lines if adaptive filters, such as Kalman filter, is employed. 
However, with the help of an accurate enough mode, Kalman filter remains a possibility to yield much superior performance than the one reported in this paper. 

Additional challenges abound in this newly proposed research domain: 
Proposed horizontal saccade tracking algorithm in this paper did not consider the the possibility of repeated reading of the same line. 
Another avenue for future research pertains to the processing of post-tracking data -- how to associate the estimated reading-progression data to latent features such as quality of experience and ``visual signature''? the answer lies in yet to be developed modeling, machine learning, and information fusion solutions.

\section*{Acknowledgements}
\label{sec:acknowledge}
B. Balasingam would like to acknowledge Natural Sciences
and Engineering Research Council of Canada (NSERC) for
financial support under the Discovery Grants (DG) program. 

\bibliography{RLS4GazeTracking}

\end{document}